\documentstyle[prl,aps,epsfig]{revtex}
\draft

\begin{document}
\twocolumn[
\hsize\textwidth\columnwidth\hsize\csname@twocolumnfalse\endcsname

\title{Zeeman smearing of the Coulomb Blockade}
\author{Karyn Le Hur}
\address{D\'epartement de Physique Th\'eorique, Universit\'e de Gen\`eve,
CH-1211, Gen\`eve 4, Switzerland.}
\maketitle
\begin{abstract}
Charge fluctuations of a {\it large} quantum dot coupled to a 
two-dimensional
lead via a single-mode {\it good} Quantum Point Contact (QPC) and 
capacitively coupled to
a back-gate,
are investigated in presence of a parallel magnetic field. The Zeeman term
induces an asymmetry between
transmission probabilities for the spin-up and spin-down channels at the
QPC, producing noticeable effects on the quantization of the grain
charge {\bf already} at low magnetic fields.
Performing a quantitative analysis, I show that 
the capacitance between the gate
and the lead exhibits --- instead of a logarithmic singularity --- a reduced
peak as a function of gate voltage. Experimental applicability is discussed. 
\end{abstract}

\vfill
 \twocolumn
\vskip.5pc ]
\narrowtext

Recent research on mesoscopic quantum dots has led
to a revival of Kondo physics. 
There is an extensive literature on the (one-channel) Kondo behavior
of small dots with an odd number of electrons 
and a finite level spacing $\epsilon$, attached to two leads. In such
an arrangement, the dot acts as an Anderson impurity\cite{Glaz-K}. 
In addition, it was shown by Matveev\cite{Matveev} 
that the Hamiltonian of a large
dot ($\epsilon\rightarrow 0$) in the Coulomb blockade
regime, coupled to a two-dimensional electron gas (2DEG)
via a good single-mode
Quantum Point Contact (QPC) and capacitively coupled to a back-gate,
is mathematically equivalent to the two-channel 
Kondo model\cite{Noz-Bla} in the Emery-Kivelson limit\cite{EK}
(Fig. 1). The 
two channels of the Kondo problem correspond to the two spin channels for
tunneling through the QPC.
For recent reviews, see Refs.\cite{Go_Ners_Tsve,ABG}. 

The non-Fermi-liquid nature of the ground state
of the two-channel Kondo model is here reflected by the non-analyticity
of the capacitance measured between the gate and the reservoir
near the points where the dot charge Q is half-integer\cite{Matveev}
$(C=\partial Q/\partial V_G)$
\begin{equation}
\label{sing}
C(N)=C_{o}-bC_{gd}|{\lambda}|^2\cos(2\pi N)\ln\frac{1}{|{\lambda}|^2
\cos^2(\pi N)}\cdot
\end{equation}
N is a parameter proportional to the gate voltage $eN=V_GC_{gd}$, $C_{gd}$
denotes the gate-dot capacitance, $C_{o}$ is the total dot capacitance
 and $|{\lambda}|^2\ll 1$ is 
the small reflection probability at the QPC. Here, $b>0$ is 
proportional to the Euler constant
$\gamma=e^{\bf C}$, with ${\bf C}\approx 0.5772$. 
The second term describes
the cross-over from the linear charge-voltage dependence to the
``Coulomb staircase'' behavior (inset in Fig. 3). This effect
has been observed by Berman {\it et al.}, in AlGaAs/GaAs 
heterostructures\cite{Dot-GaAs,Matveev2}.

Below, I investigate how 
the capacitance $C(N)$ evolves
if a magnetic field B is applied parallel to the 2DEG\cite{note}. 
I will show that the logarithmic
divergence in the differential
capacitance $\delta C(N)=C(N)-C_o$ with $N\rightarrow 1/2$
gets already cutoff by a {\it small} magnetic field (Fig. 3)
\begin{eqnarray}
\label{deux}
\frac{\delta C}{C_{gd}}(N)&=& b|{\lambda}|^2(1-{\hat{\delta}}^2)\cos(2\pi N)
\times
\\ \nonumber
& &
\ln{\hbox{Max}}\hbox{\Large{[}}|{\lambda}|^2\hat{\delta}^2\sin^2(\pi N);
|{\lambda}|^2\cos^2
(\pi N)\hbox{\Large{]}},
\end{eqnarray}
$0<\hat{\delta}\ll 1$, being proportional to the Zeeman energy, 
$\Delta=g\mu_B B$. This results from a Zeeman-like asymmetry between
reflection probabilities for the spin-up and spin-down channels at the QPC.
For a high field $(\hat{\delta}\approx 1)$, 
only the spin-up channel will be transmitted producing a
one-channel QPC model. The capacitance (or charge Q)
only exhibits periodic oscillations as a function of N\cite{Matveev}, 
\begin{equation}
\label{capa}
C(N)=C_{o}-bC_{gd}|{\lambda}|\cos(2\pi N).
\end{equation} 
The Coulomb staircase behavior gets completely smeared out (inset in Fig. 3).
An experiment using AlGaAs/GaAs heterostructures
is presumably appropriate to probe this 
effect in the capacitance\cite{expe-rem}. Indeed, Zeeman splitting at 
the QPC
in an in-plane magnetic field 
has been confirmed by Thomas {\it et al} via conductance 
measurements\cite{Thomas}, and for few conducting modes at the 
QPC the Land\'e factor is enhanced $g\approx 1$\cite{Patel}
(the bulk value is $|g|=0.4$).

Let me emphasize that below a {\it quantitative} analysis of the 
smearing out of the logarithmic peak
for the capacitance $\delta C(N)$ is performed (which has not been
previously done in Ref.\cite{Matveev}). The crossover from the two-
to the one-channel QPC model is carefully investigated.
\vskip 0.1cm
First, it is useful to compute reflection probabilities at the QPC 
in presence of a magnetic field parallel to the
2DEG, and to discuss the necessary magnetic-field dependent adjustment
of the QPC potential $V_a(B)$. In the close vicinity of the contact, the
(smooth) confining potential will be approximated as a 
harmonic one\cite{Go_Ners_Tsve}.
\vskip -0.1cm
\begin{figure}[ht]
\centerline{\epsfig{file=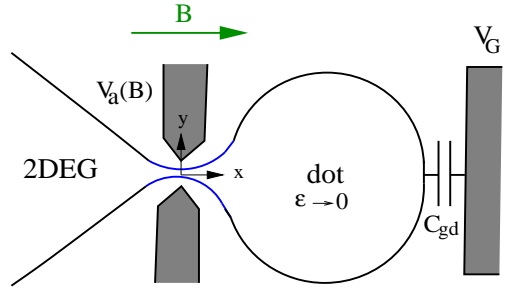,angle=0.0,height=3.8cm,width=6.6cm}}
\vskip 0.3cm
\caption{The experimental setup: A large dot coupled to
a 2DEG via a single mode QPC and capacitively coupled to a back-gate, 
in a parallel magnetic field B ($C_o=C_{gd}$).}
\end{figure}

An electron with spin-projection $\alpha=\uparrow,\downarrow$ 
(or $\alpha=\pm 1$) along the magnetic field axis, is then subjected to the 
total potential
$V_a(B)-m{\omega_x}^2x^2/2+m{\omega_y}^2y^2/2-\alpha\Delta/2$. Here, 
$\omega_x$ and
$\omega_y$ are the curvatures of the potential, and $m$ the 
mass of an electron. 
The transverse part of the Hamiltonian produces ``n transverse 
modes''. Then, the one-dimensional 
(1D) wave function ${\Psi}^n_{\alpha}(x)$ for motion along $x$ 
is determined by the effective potential
\begin{equation}
{V}_n^{\alpha}(x)=V_a(B)-\alpha\frac{\Delta}{2}
+\hbar\omega_y(n+\frac{1}{2})-\frac{1}{2}m{\omega_x}^2 x^2,
\end{equation}
the explicit form of which is of no interest here.
The height of the barrier potential at the QPC for a
spin-down electron is enhanced by 
$\Delta$. From now on, I adjust $V_a(B)=V_a(B=0)+\Delta/2$, because I want  
field-independent reflection probabilities for spin-up channels.

Near the saddle-point $(x\approx 0)$, 
the threshold energies of the mode $n$ are then spin-splitted, as follows
\begin{eqnarray}
\label{Zeeman}
E_n^{\uparrow}&=&E_n=V_a(B=0)+\hbar\omega_y(n+\frac{1}{2}),
\\ \nonumber
E_n^{\downarrow}&=& E_n+\Delta.
\end{eqnarray} 
Classically, modes with threshold energy below the Fermi energy $E_F$ are 
perfectly open and the others remain closed. But, quantum mechanically
transmission and reflection at the saddle are neither completely open nor
completely closed\cite{Markus-cond}. Here, I fix the voltage $V_a(B=0)$ 
such that $E_0\ll E_F< E_1$. In Fig. 2, this 
corresponds to adjust $\xi$ such that $0.5\ll\xi< 1.5$ 
$(\omega_y/\omega_x=1)$. Then, the spin-up channel of the transverse 
mode n=0 remains at almost perfect 
transmission whatever the applied magnetic field. Moreover, I can 
disregard modes $n\geq 1$ which
are almost perfectly reflected, because I am interested only in  
transport through the constriction.

The reflection/transmission amplitude of a 1D particle passing through
an inverted parabolic barrier has been studied in detail by 
Connor\cite{Connor}.
Taking $n=0$, small reflection probability for the spin-up channel reads
\begin{equation}
\label{up}
{\cal R}_0^{\uparrow}=|\lambda|^2=\frac{1}{1+\exp(2\pi{\cal E}^{\uparrow}_0)}
\approx\exp(-2\pi{\cal E}^{\uparrow}_0),
\end{equation}
where $
{\cal E}^{\uparrow}_0=[E_F-E_0^{\uparrow}]/\hbar\omega_x\gg 0$.
 Below, I will focus on low magnetic
field effects. Using Eq. (\ref{Zeeman}), similarly I obtain
$(\delta=\Delta/2\hbar\omega_x\ll 0.4)$
\begin{equation}
\label{down}
{\cal R}_0^{\downarrow}\approx\exp(-2\pi{\cal E}^{\downarrow}_0)\approx
{\cal R}_0^{\uparrow}(1+4\pi\delta).
\end{equation}
Both channels are transmitted
but $1\gg{\cal R}_0^{\downarrow}>{\cal R}_0^{\uparrow}$ (See
Fig. 2). Applying
an in-plane magnetic field, 
one gets what I call a two-channel anisotropic QPC model; This
is defined, below. The limit of 
strong fields will be reached when the Zeeman energy
approaches the curvature energies of the potential:
$E_0^{\downarrow}=E_1$ and ${\cal T}_0^{\downarrow}=1-{\cal R}_0^{\downarrow}\approx 0$. A single 
channel will subsist in the constriction.
\begin{figure}[ht]
\centerline{\epsfig{file=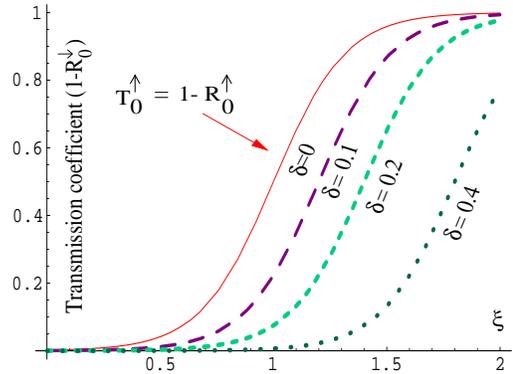,angle=0.0,height=5cm,width=6.7cm}}
\vskip 0.3cm
\caption{Exact transmission probability for the transverse
mode $n=0$ and the spin-channel opposite to the
magnetic field, as a function of 
$\xi=[E_F-V_a(B=0)]/\hbar\omega_x$. The
ratio $\omega_y/\omega_x$, being 1. The
different curves correspond to different values of the magnetic field i.e.
$\delta=\Delta/2\hbar\omega_x=0,0.1,0.2,0.4$.}
\end{figure}

As I am interested in the dynamics of the system at energies much smaller
than $E_F$, I may linearize the spectrum of the 1D-fermions in state 
$\Psi_{\alpha}(x)=\Psi^{n=0}_{\alpha}(x)$. One can always
write $\Psi_{\alpha}(x)=
\exp(ik_F x)\Psi_{R\alpha}(x)+\exp(-ik_F x)\Psi_{L\alpha}(x)$; 
$\Psi_{L\alpha}$ and $\Psi_{R\alpha}$ respectively describe
left- and right moving fermions, and $E_F={k_F}^2/2m$. Finite reflection 
in the channel $\alpha=\uparrow,\downarrow$ then
can be simply accounted for by adding a backscattering 
term\cite{Go_Ners_Tsve,ABG,Aleiner}
\begin{equation}
\label{backs}
H_{bs}=
v_F\sqrt{{\cal R}_0^{\alpha}}\Psi_{L\alpha}^{\dagger}(0)\Psi_{R\alpha}(0).
\end{equation} 
$v_F$ denotes the Fermi velocity\cite{vel}. 
One must also include the kinetic energy through the constriction,
\begin{equation}
H_{kin}=i v_F\int_{-\infty}^{+\infty}dx \hbox{\Large{\{}} 
\Psi_{R\alpha}^{\dagger}\partial_x\Psi_{R\alpha}-
\Psi_{L\alpha}^{\dagger}\partial_x\Psi_{L\alpha}
\hbox{\Large{\}}}.
\end{equation}
At almost perfect transmission, the electronic wave function is shared
between the reservoir and the dot. I can neglect finite
size effects in a dot at the micron scale
$(\epsilon\rightarrow 0)$\cite{ABG,Aleiner,Markus}.
Note that $(H_{kin}+H_{bs})$ with
$1\gg{\cal R}_0^{\downarrow}>{\cal R}_0^{\uparrow}$
describes a {\it two-channel anisotropic QPC model}. 
Again, I ignore higher modes
confined to the reservoir, and also neglect 
the Pauli contribution of the 2DEG. The charging process is 
described by the following usual term
\begin{equation}
H_c=E_c\hbox{\Large{(}}Q-N\hbox{\Large{)}}^2,
\end{equation}
with $E_c=e^2/(2C_{gd})\ll E_F$ the (charging) energy that it
costs to transfer a particle from the lead to the dot.
The charge Q (of the dot) in $H_c$ is {\it now} normalized to e,
\begin{equation}
Q=\int_0^{+\infty}dx 
\hbox{\Large{\{}} \Psi_{L\alpha}^{\dagger}\Psi_{L\alpha}+
\Psi_{R\alpha}^{\dagger}\Psi_{R\alpha}\hbox{\Large{\}}}.
\end{equation}

At low energies, I can proceed with this model by bosonizing the 
1D Fermi fields\cite{Go_Ners_Tsve,ABG,Aleiner,KLH},
\begin{equation}
\label{boso}
\Psi_{p\alpha}=\frac{1}{\sqrt{2\pi a}}
\exp\hbox{\Large(}i\sqrt{\frac{\pi}{2}}\left[p(\phi_{c}+\alpha\phi_{s})
    -(\theta_{c}+\alpha\theta_{s})\right]\hbox{\Large )}.
\end{equation}
$a$ is a short-distance cutoff, again
$\alpha=\pm$ for spin up and spin down, and $p=\pm$ for right
and left movers. The spectrum of 1D free electrons yields separation
of spin and charge. Resulting Hamiltonians are plasmon-like
\begin{equation}
H_{kin}=\sum_{j=c,s}\frac{v_F}{2}
\int_{-\infty}^{+\infty}\ dx\ \hbox{\Large{[}}{(\partial_x\phi_j)}^2+{\Pi_j}^2
\hbox{\Large{]}}.
\end{equation}
$\partial_x\phi_j$ with $j=(c,s)$ measures fluctuations of charge/spin density
in the constriction and $\Pi_j=\partial_x\theta_j$ 
being its conjugate momentum. In this 
representation, the charging Hamiltonian $H_c$ reads
\cite{Go_Ners_Tsve,ABG,Aleiner}
\begin{equation}
H_c=E_c\hbox{\Large{[}} \sqrt{\frac{2}{\pi}}\phi_c(0)-N\hbox{\Large]}^2.
\end{equation}
To minimize energy, the charge in the dot is pinned at the 
{\it classical} value
$Q_c=\phi_c(0)\sqrt{2/\pi}\approx N$\cite{Go_Ners_Tsve,ABG,cb}. 

Now, one has to examine the
quantum corrections to the charge entering the constriction, in
presence of a small (parallel) magnetic field.
For energies smaller than $E_c$ and $|\lambda|\ll 1$, I can
replace $\cos[\sqrt{2\pi}\phi_c(0)]$ by the averaged value
$\sqrt{\gamma E_c a/\pi v_F}\cos(\pi N)$ (similarly
for the sinus). The prefactor
comes from the average of the cosine term over the ground state of 
$H_{kin}$\cite{Glazman,Aleiner}. 
Using Eqs. (\ref{up}),(\ref{down}) and (\ref{boso}) the
backscattering term then reads
\begin{eqnarray}
\label{back}
H_{bs}&\approx&\frac{|\lambda|}{2\pi a}\sqrt{\gamma v_F E_c a}\cos(\pi N)
\cos[\sqrt{4\pi}\Phi_s(0)]\\ \nonumber
&+&
\frac{|\lambda|\hat{\delta}}{2\pi a}\sqrt{\gamma v_F E_c a}\sin(\pi N)
\sin[\sqrt{4\pi}\Phi_s(0)].
\end{eqnarray} 
{\it The previous anisotropy parameter
$\hat{\delta}\ll 1$ is equal to $\pi\delta$}.
As in Ref.\cite{Go_Ners_Tsve} (page 417), I 
introduce the symmetric combination of the spin
Bose fields with respect to the QPC, 
$\phi_s(x)=[\Phi_s(x)+\Phi_s(-x)]/\sqrt{2}$. The kinetic part is\cite{rem}
\begin{equation}
H_{kin}[\Phi_{s}]=v_F\int_{-\infty}^{+\infty}
\ dx\ {(\partial_x\Phi_s)}^2.
\end{equation}

In the absence of the magnetic field, i.e. $\hat{\delta}=0$, the Hamiltonian 
$H_{kin}[\Phi_{s}]+H_{bs}$ is equivalent to the two-channel
Kondo model at the Emery-Kivelson line
\begin{equation}
H_{EK}=H_{kin}[\Phi_{s}]+\frac{J_{\perp x}[N]}{\pi a}
\cos[\sqrt{4\pi}\Phi_s(0)]\hat{S}_x,
\end{equation}
with the Kondo exchange $J_{\perp x}=|\lambda|\sqrt{\gamma 
v_F E_c a}\cos(\pi N)$. $\hat{S}_x$
describes the x-component of a fictitious quantum impurity.
This equivalence occurs because $\hat{S}_x$ commutes with 
$H_{EK}$\cite{Go_Ners_Tsve}. One can set $\hat{S}_x=1/2$ 
(or $\hat{S}_x=-1/2$\cite{notes}).
$H_{EK}$ can be exactly 
refermionized and diagonalized\cite{EK,Go_Ners_Tsve}.
The backscattering contribution to the ground state energy
then takes the form
$\delta E(N)=-\frac{\Gamma(N)}{2\pi}\ln(E_c/\Gamma(N))$, where $\Gamma(N)=
{J_{\perp x}}^2/\pi v_F a$ is the Kondo resonance. The quantum
correction $\delta Q$ to the charge in the dot, becoming equal to
$\delta Q=-\partial\delta E/(E_c\partial N)$\cite{Matveev,Go_Ners_Tsve}. 
This results explicitly in 
$\delta Q\approx -b{\cal R}_0^{\uparrow}\sin(2\pi N)\ln(E_c/\Gamma(N))$\cite{Matveev}.
\begin{figure}
\centerline{\epsfig{file=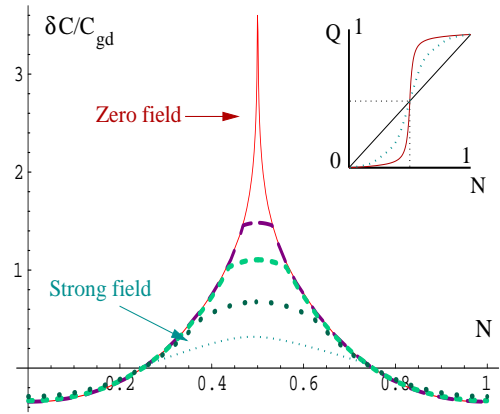,angle=0.0,height=5.5cm,width=6.5cm}}
\vskip 0.3cm
\caption{Normalized differential gate-lead capacitance as a function of N, for different
 values of the anisotropy parameter $\hat{\delta}=\pi\delta=0,0.1,0.2,0.4,1$
(from top to bottom). In Fig. 2, I fix $\xi=1.15$ such
that ${\cal R}_0^{\uparrow}=0.25$, and $b=1$. I also draw the total charge
$Q=Q_c+\delta Q$ as a function of N for $\hat{\delta}=0,1$.}
\end{figure}

The charge fluctuations get ``strongly'' influenced by the spin
ones; The logarithmic factor in $\delta Q$ 
comes from spin fluctuations at the QPC\cite{Matveev,Go_Ners_Tsve}. For 
$\hat{\delta}=0$, these get pinned at the time $\hbar/\Gamma$ at which the 
screening of $\hat{S}_x$ takes place and the Kondo
coupling $J_{\perp x}$ diverges. For N=1/2, $\Gamma=0$: There
is a degeneracy between the dot states with Q=0 and Q=1 [See
Fig. 3; $Q=Q_c+\delta Q$]. The capacitance displays a logarithmic divergence at this degeneracy point [Eq. (\ref{sing})]. If $N<1/2$ one gets
$\sqrt{4\pi}\Phi_s(0)\approx \pi$, and 
$\sqrt{4\pi}\Phi_s(0)\approx 2\pi$ if $N>1/2$\cite{notess}. 
For $\hat{\delta}\neq 0$ the backscattering contributions to 
the ground state energy 
can now be calculated using the two-channel anisotropic Kondo model at 
the Emery-Kivelson line\cite{Fabrizio}
\begin{equation}
H_{EK}^{A}=H_{EK}-
\frac{J_{\perp y}[N]}{\pi a}\sin[\sqrt{4\pi}\Phi_s(0)]\hat{S}_y.
\end{equation}
The extra magnetic exchange is (See Ref.\cite{Go_Ners_Tsve}, page
394)
\begin{equation}
\label{J}
J_{\perp y}[N]=|\lambda|\hat{\delta}\sqrt{\gamma v_F E_c a}\sin(\pi N),
\end{equation} 
$\hat{S}_y$ being the y-component of the fictitious 
impurity.
\vskip 0.15cm
The equivalence between $H_{kin}[\Phi_s]+H_{bs}$ and
$H_{EK}^{A}$ can be understood, as follows. Starting with 
low-magnetic fields and at almost
perfect transmission, the following commutators can be neglected whatever
$N\in[0;1]$ (for frequencies larger than $\omega_{min}$; See 
discussion below)
\begin{eqnarray}
[H_{EK}^{A},\hat{S}_x] &\propto& i\hat{\delta}|\lambda|\hat{S}_z\ll 1
\\ \nonumber
[H_{EK}^{A},\hat{S}_y] &\propto& i|\lambda|\hat{S}_z\ll 1.
\end{eqnarray}
This implies that $\hat{S}_x$ and $\hat{S}_y$ both can be considered
as good quantum numbers: The impurity spin in $H_{EK}^{A}$ can 
oscillate between the two states $\hat{S}_x=1/2$ and $\hat{S}_y=-1/2$.
The use of the 
two-channel anisotropic Kondo model is completely justified, for all 
$N\in[0;1]$. It is noteworthy that this model can be refermionized
and diagonalized; See e.g. Ref.\cite{Go_Ners_Tsve}, page 401. 
Another energy scale related to the magnetic field emerges:
$\Upsilon={J_{\perp y}}^2/\pi v_F a=\gamma E_c
{\cal R}_0^{\uparrow}{\hat{\delta}}^2\sin^2(\pi N)/\pi$. This is associated
with the divergence of $J_{\perp y}$. Using the two-channel anisotropic
Kondo model, for $B\neq 0$ I find $\hbox{\Large{(}}\Im m\{
\ln[i\Gamma-\omega]\}=-\tan^{-1}[\Gamma/\omega]\hbox{\Large{)}}$
\begin{equation}
\label{E}
\delta E = +\int^{E_c/\hbar}_{{\omega}_{min}} 
\frac{d\omega}{2\pi} \Im m\hbox{\Large{\{}}\ln[i\Gamma-\omega]
+\ln[i\Upsilon-\omega]\hbox{\Large{\}}}.
\end{equation}
Let me comment on the low-frequency (spin) cutoff. 

Away from half-integer values of N, 
$\Upsilon$ tends to zero because $\sin(\pi N)\ll 1$. Then,
the coupling $J_{\perp y}\ll 1$ is not really affected by the 
renormalization of the
short-distance cutoff $a$. As in the absence of the magnetic field, all 
spin fluctuations get pinned at the time $\hbar/\Gamma$.
The equilibrium state at the QPC 
then corresponds to $\sqrt{4\pi}\Phi_s(0)\approx \pi\ [2\pi]$ if 
$N\ll 1/2$ [if
$N\gg 1/2$], and $J_{\perp y}\hat{S}_y<\sin[\sqrt{4\pi}\Phi_s(0)]>\ 
\approx 0$. Approaching the point $N=1/2$, in contrast the Kondo
resonance vanishes. But now, $\Upsilon$ becomes dominant such that
$J_{\perp y}$ is flowing off to
strong couplings at the {\it finite} time $\hbar/\Upsilon$. Spin 
fluctuations at the
QPC then get pinned also for $N\approx 1/2$ and
$\sqrt{4\pi}\Phi_s(0)\approx 3\pi/2$. 
In the presence of the magnetic field, therefore the
logarithmic integral on spin fluctuations in Eq. (\ref{E}) now
acquires a natural low-energy cut-off, 
$\hbar\omega_{min}=\hbox{Max}[\Gamma;\Upsilon]$,
whatever the value of $N\in[0;1]$.
Performing the integration in Eq. (\ref{E}), then I precisely
obtain
$\delta Q\approx -b{\cal R}_0^{\uparrow}(1-{\hat{\delta}}^2)\sin(2\pi N)\ln(E_c/\hbar\omega_{min})$. 

The charge (and spin) in 
the dot becomes a ``continuous''
function of N. 

This enables me to allege that the 
capacitance $C=\partial Q/\partial V_G$ displays, instead of a 
logarithmic singularity, a reduced
peak as a function of N; See Eq. (\ref{deux}) and Fig. 3. For a small 
deviation $\hat{n}=(N-1/2)$ from $N=1/2$, one 
gets $\Gamma(\hat{n})\propto {\hat{n}}^2$ and 
$\Upsilon(\hat{\delta})\propto {\hat{\delta}}^2$, and then
\begin{equation}
\delta C \propto -\left\{
\begin{array}{r@{\quad:\quad}l}
\ln\hat{\delta} & \hat{n}\ll\hat{\delta}\\ 
\ln\hat{n} & \hat{\delta}\ll\hat{n}.
\end{array} \right. 
\end{equation}
Increasing the in-plane magnetic field, i.e. $\hat{\delta}$, then the 
capacitance peak reduces and becomes more and more broadened. For high fields 
$\hat{\delta}\approx 1$ and $\delta\approx 0.4$,
only spin-up electrons can be transmitted:
${\cal T}_0^{\downarrow}\rightarrow 0$. One must disregard the spin-down
channel at the QPC. Formally, $\hbar\omega_{min}$ now 
exceeds $E_c$ and the classical
value of the grain charge must be rescaled as:
$Q_{c}=\phi_{\uparrow}(0)/\sqrt{\pi}
=[\phi_c(0)+\phi_s(0)]/\sqrt{2\pi}\approx N$. Now, this produces
$\delta E=-|\lambda| \gamma E_c\cos(2\pi N)/{\pi}^2$ and then
only a weak residual oscillation in
$\delta C=-b|\lambda|\cos(2\pi N)$ for all $N\in[0;1]$, as for 
``spinless'' fermions\cite{Matveev}. The logarithmic part 
disappears.

To conclude, investigating zero-temperature
properties of a large dot coupled to a two-channel anisotropic QPC, 
I have shown that the steps of the Coulomb staircase function gets 
immediately smeared out by a finite magnetic field parallel to
the 2DEG. This rigorously proves that like the magnetic susceptibility
of the Kondo impurity in the two-channel Kondo model\cite{Fabrizio}, the 
logarithmic singularity in the capacitance
is destroyed by a small channel anisotropy.
For high fields, adjusting correctly the voltage at the QPC one
reproduces a single-channel QPC model; For $|\lambda|\ll 1$,
the charge entering the dot
exhibits only a weak quantum oscillation around its 
classical value. Experimentally, the 
prominent smearing 
of the Coulomb staircase --- predicted by increasing the in-plane 
magnetic field ---
could be possibly observed for sufficiently low temperatures T, probing 
(preferably) the capacitance line shapes
of the large 
dot in a single-terminal geometry\cite{expe-rem}. To account for 
nonzero temperature, 
one must rescale $\hbox{Max}[\Gamma;\Upsilon]$ as 
$\hbox{k}_B\hbox{T}+\hbox{Max}[\Gamma;\Upsilon]$. Experimentally, it 
would be
crucial to minimize the broadening of the capacitance peaks due to 
thermal effects.
\vskip 0.05cm
This work is supported by the Swiss National Science Foundation.


\begin{references}
\bibitem{note} An in-plane magnetic field is judicious to probe 
spin-splitting effects because the Lorentz force $-e\vec{v}\wedge\vec{B}
\rightarrow 0$, that 
allows to completely eliminate the cyclotron energy.
\bibitem{expe-rem} In principle, to reproduce the result of Eq.(2) 
would require 
a smooth confining potential at the QPC and a ``perfect'' dot with no
dephasing/relaxation process.
\bibitem{vel} At low magnetic fields $\Delta\ll E_F$, one can neglect the 
renormalization of the Fermi velocities for spin-up and spin-down electrons.
This does not affect the final result.
\bibitem{notes} The spectrum is indeed a function of ${J_{\perp x}}^2$.
\bibitem{notess} For $N<1/2$ ($N>1/2$), the corresponding
spin state in the dot is S=0 (S=1/2).
\end{references}
\end{document}